\documentclass[aps,pre,twocolumn,showpacs]{revtex4-1}
\usepackage{amsmath,amssymb}
\usepackage{graphicx}
\usepackage[latin1]{inputenc}
\usepackage{bm}  

\usepackage{color,soul}
\setstcolor{red}

\newcommand{\be}{\begin{equation}}
\newcommand{\ee}{\end{equation}}
\newcommand{\ba}{\begin{eqnarray}}
\newcommand{\ea}{\end{eqnarray}}

\def\LM#1#2{\left|\begin{array}{l}{#1}\\[1ex]{#2}\end{array}\right.}

\begin{document}

\title{Survival probability of an immobile target
in a sea of evanescent diffusive or subdiffusive traps: a fractional equation approach}
\author{E. Abad$^{1}$, S. B. Yuste$^{2}$ and Katja Lindenberg$^{3}$}
\affiliation{$^{(1)}$
Departamento de F\'{\i}sica Aplicada, Centro Universitario de Mérida, Universidad de Extremadura,
E-06800 Mérida, Spain
\\
$^{(2)}$
Departamento de F\'{\i}sica, Universidad de Extremadura,
E-06071 Badajoz, Spain
\\
$^{(3)}$ Department of Chemistry and Biochemistry, and BioCircuits Institute,
University of California San Diego, 9500 Gilman Drive, La Jolla, CA
92093-0340, USA
}

\date{\today}

\begin{abstract}

We calculate the survival probability of an immobile target surrounded by a sea of
uncorrelated diffusive or subdiffusive evanescent traps, i.e., traps that disappear
in the course of their motion.  Our calculation is based on a fractional
reaction-subdiffusion equation derived from a continuous time random walk model of the system.
Contrary to an earlier method valid only in one dimension ($d=1$), the equation is applicable in any Euclidean
dimension $d$ and  elucidates the interplay between anomalous subdiffusive transport, the irreversible evanescence reaction and the dimension in which both the traps and the target are embedded.
Explicit results for the survival probability of
the target are obtained for a density $\rho(t)$ of traps which decays \textit{(i)} exponentially and
\textit{(ii)} as a power law. In the former case, the target has a finite asymptotic survival
probability in all integer dimensions, whereas in the latter case there are several regimes
where the values of the decay exponent for $\rho(t)$ and the anomalous diffusion exponent of
the traps determine whether or not the target has a chance of eternal survival in one, two and
three dimensions.

\end{abstract}


\maketitle

\section{Introduction}

Geometric and dynamical constraints imposed by complex or crowded environments often result in
subdiffusive behavior, i.e., in sublinear growth of a particle's mean squared displacement at long times. However, a complete description of the underlying transport process at a mesoscopic level must go beyond the mean squared displacement and involve other properties of experimental interest which may be studied via suitable quantifiers \cite{SokolovReview2, Magdziarz}. This may help one to discriminate between models when describing realistic experimental situations where subdiffusive (or, more generally, anomalous) transport is observed.

The detailed microscopic subdiffusive transport mechanism is often unknown, and so the literature is populated with a number of different models. One popular choice to mimic situations of experimental interest is the continuous time random walk (CTRW) model \cite{MontrollWeiss} with a long-tailed waiting time distribution. The CTRW has been used successfully as a phenomenological model to describe aging effects in systems as diverse as stock markets \cite{ScalasGorenfloMainardiPAI, ScalasGorenfloMainardiPAII, Masoliver1, Masoliver2}, charge carrier transport in disordered media \cite{ScherLaxI}, luminescence quenching in micellar clusters \cite{BarzykinTachiyaPRL}, transport in porous media \cite{Berko, LeBorgneEtAl}, escape problems \cite{divergentSeriesYBA}, and morphogen gradient formation \cite{HornungPRE06, YusteAbadLindenberg10, BoonLutsko}.

From a mathematical point of view, the CTRW with a long-tailed waiting time distribution and a jump length distribution of finite variance is known to be equivalent to a fractional diffusion equation in the long-time limit, that is, a diffusion equation with fractional time derivatives rather than ordinary derivatives  \cite{MetzlerKlafterPhysRep}. Despite the fact that fractional derivatives are non-local integro-differential operators, Laplace transform techniques commonly used for the solution of the ordinary diffusion equation remain applicable and can be used to tackle a wide variety of these problems \cite{BookChapter}.

One compelling reason to work with CTRW models and the associated fractional equations is that they make it possible to include reactive processes. We introduce this terminology in the broadest sense of including particle destruction, creation, binding or transformation processes. While the combination of subdiffusion with its memory effects and reaction processes is complex, at least the CTRW approach offers a way to consider them in combination, something that has proved more elusive with other approaches. In some fortunate situations, the effect of the reactions can be adequately described by suitable boundary conditions imposed upon the corresponding fractional diffusion equation (see e.g. \cite{BorregoAbadYuste}, Section 4.1 in \cite{BookChapter}, and references therein); however, such situations are rather exceptional, since in general the combination of reaction with non-Markovian kinetics \cite{MFHbook} leads to non-intuitive fractional equations where the parameters describing the chemical kinetics appear in a non-universal, model-dependent fashion \cite{HenryWearne00, SokolovSchmidtSaguesPRE06, Henry, Yadav06, AbadYusteLindenberg10} . In particular, heuristic approaches based on  fractional equations with separate reaction and transport terms such as we are accustomed to in ordinary reaction-diffusion problems very often lead to unphysical results even in the simplest cases of irreversible first-order reactions.

While fractional reaction-subdiffusion equations have been used to investigate a number of different problems corresponding to different mesoscopic models and different boundary conditions \cite{SekiWojcikTachiyaJCP03, SekiShushinWojcikTachiyaCM07, FedotovIomin2007, Yadav08, FroembergSokolovPRL08, YusteAbadLindenberg10, GafiychukDatsko}, many subdiffusive versions of classical reaction-diffusion problems \cite{CrankBook} remain unexplored. Thus, one can legitimately claim that the field is still in its infancy. One class of problems that has attracted considerable interest in recent years concerns target search processes driven by (sub)diffusion. Such processes are ubiquitous in nature and include binary searches where two objects must meet for a reaction or trapping event to occur. In many instances, Smoluchowski's theory of diffusion controlled reactions turns out to be a successful tool for the quantitative characterization of diffusional target search. Examples include scavenging reactions \cite{SanoTachiyaJCP79, KimEtAl}, site location in DNA \cite{VonHippelBerg}, ligand binding to sites on macromolecules \cite{BerezhkovskiiEtAlJCP11}, predator-prey models \cite{RednerKrapivsky},  luminescence quenching \cite{Tachiya}, intermittent search processes \cite{BenichouEtAlRMP11}, and search processes with resetting to the initial position \cite{EvansMajumdar}, to name but a few. In this context a key quantity is the so-called \emph{survival probability} of the target, from which the moments of the first-passage-time distribution for target annihilation can also be straightforwardly computed \cite{Weiss-Book, CondaminBenichouMoreau07}.

In recent years the classic diffusional target search problem has been generalized
to particles that undergo anomalous diffusion \cite{targetyuste, EavesReichman, Grebenkov, FrankeMajumdar}. In this paper we consider a related problem, namely, the survival probability of an immobile target immersed in a sea of uncorrelated subdiffusive traps \emph{that may die ``spontaneously'' in the course of their motion}. In other words, there are now two reactions occurring simultaneously: the disappearance of the target and a trap upon encounter with each other, and the disappearance of the traps due to some \emph{other} physical process. We term this latter process ``spontaneous'' as a way to recall that it is not induced by collision with the target. The spontaneous evanescence process may for instance be triggered by particle scavengers in the system, but for practical purposes any process that turns off the interaction between a trap and the target can also be thought of as an evanescence or death process.

A solution to this problem in dimension $d=1$ was given in Ref.~\cite{YRLevanescenciaPRE}
using a functional method first developed by Bray et al.~\cite{Brayetal} for the diffusive case. Here
we approach the problem from a different point of view that allows us to also obtain results in higher dimensions.  In particular, we make use of a recently
derived reaction-subdiffusion equation obtained from a mesoscopic CTRW model with a long-tailed
waiting time distribution and a superimposed reactive process. The elimination of the fixed target is incorporated as a boundary condition, while the decay mechanism of the traps as they move subdiffusively is modeled by a phenomenological choice of a monotonically decaying functional form for the trap density $\rho(t)$.

The paper is organized as follows. In Sec.~\ref{connevnev} we briefly recall the results for the survival probability of a target in a sea of non-evanescent traps. Our new general results for the survival probability of the
target when the traps are evanescent are presented in Sec.~\ref{genexpressions}. In Sec.~\ref{implementation}
we implement these results for particular forms of evanescence, namely, exponential and power law.
We conclude with a short summary in Sec.~\ref{summary}.

\section{Recap of results for non-evanescent traps}
\label{connevnev}

We consider a statistical ensemble of systems each of which is composed of
a fixed hyperspherical target of radius $R$ located at the center $r=0$ of a large $d$-dimensional
volume $V$.  Each volume $V$ initially contains  $N_0$
randomly distributed non-interacting point traps. At time $t=0$ the traps begin to move subdiffusively. If any
of them hits the surface of the target, both the target and the trap are instantaneously annihilated
(fully absorbing case). Our goal is to compute the survival probability of the target at
time $t$, i.e., the probability that no traps have collided with the target up to this time.
In this section, we briefly recall previous results obtained in Ref.~\cite{targetyuste} when
the traps are not subject to spontaneous evanescence. In the next section we use these
results to obtain the solution when the subdiffusive traps evanesce.

In the absence of evanescence, the motion of each trap is dictated by the fractional diffusion equation
\begin{equation}
\label{fde}
\frac{\partial w(\mathbf{r},t|\mathbf{r}_0;0)}
{\partial t}=K_{\gamma}
 \,{_{0}\cal D}_t^{1-\gamma} \, \nabla_{\mathbf{r}}^2\,w(\mathbf{r},t|\mathbf{r}_0;0), \qquad 0<\gamma \le 1,
\end{equation}
where $w(\mathbf{r},t|\mathbf{r}_0;0) $ is the probability density of finding the trap at location $\mathbf{r}$ at time $t$ if it started at position $\mathbf{r}_0$ at $t=0$, $K_\gamma$
is the anomalous diffusion coefficient, and $\nabla_{\mathbf{r}}^2$  stands for  the Laplacian
operator with respect to the position $\mathbf{r}$. The operator ${_{0} \cal D}_t^{1-\gamma}$ is defined via the equation
\begin{equation}
{\cal L}_{t\to u}\left\{{_{0} \cal D}_t^{1-\gamma}f(t)\right\}=u^{1-\gamma}{\cal L}_{t\to u}
\left\{f(t)\right\},
\end{equation}
where ${\cal L}_{t\to u}\{f(t)\}\equiv {f}(u)=\int_0^\infty e^{-ut}f(t)\,dt$ denotes the
Laplace transform (the function and its Laplace transform are clearly distinguished by the argument and so we use the same designation for both). Instead of ${_{0} \cal D}_t^{1-\gamma}$ we employ the more commonly used
Riemann-Liouville fractional derivative $_{0}D_t^{1-\gamma}$ defined as follows~\cite{Podlubny}:
\begin{equation}
~_{0}D_t^{1-\gamma} f(r,t)
=\frac{1}{\Gamma(\gamma)}\frac{\partial}{\partial t}
\int_0^t dt'\, \frac{f(r,t')}{(t-t')^{1-\gamma}}.
\end{equation}
Strictly speaking, the two operators are not identical.  However, they
are equivalent for functions which are sufficiently regular at $t=0$, as are
all the functions we encounter in this problem.
The propagator solution of the fractional diffusion equation \eqref{fde} yields
a mean squared displacement with the long time behavior $\langle r^2 \rangle\sim K_\gamma t^\gamma$  resulting in subdiffusive behavior when $\gamma$ is less than unity.

Let $Q_T(t;R)$ denote the ensemble averaged survival probability of the target in a sea of
randomly distributed uncorrelated traps. This quantity
can be obtained from the survival probability of
the target in the presence of a \textit{single} trap starting at location $\mathbf{r}_0$, $Q_{1,T}(\mathbf{r}_0,t;R)$.
We shall focus on the thermodynamic limit, i.e., we take $N_0\to\infty,V\to \infty$ while
keeping a fixed global initial trap density $\rho_0=\lim_{N_0,V \to\infty} N_0/V$.
In this limit one has
\begin{eqnarray}
\label{Q*Ta}
  Q_T(t;R)&=&\lim_{N_0,V\to\infty} \left[\frac{1}{V} \int_{r_0>R}
  Q_{1,T}(\mathbf{r}_0,t;R)\, d\mathbf{r}_0\right]^{N_0}\nonumber\\\nonumber\\
&=&\exp\left\{-\rho_0 R^d \sigma(t,R) \right\},
\end{eqnarray}
where the integration is carried out over the volume that is exterior to the target.
We have introduced the auxiliary quantity
\begin{eqnarray}
 \sigma(t;R)&\equiv &
 \frac{1}{R^d} \int_{r_0>R}[1-Q_{1,T}(\mathbf{r}_0,t;R)]\,d\mathbf{r}_0\nonumber\\\nonumber\\
 &=& -\frac{1}{\rho_0 R^d} \ln Q_T(t;R).
 \label{sigmatR}
\end{eqnarray}
Note that the survival probability $Q_{1,T}(\mathbf{r}_0,t;R)$ of the target is identical with the
survival probability $Q_1(\mathbf{r}_0,t;R)$ of the trap, as we have assumed that both the target and the
trap disappear instantaneously upon encounter, i.e., there is only one decay channel for both
particles. Note also that this is no longer the case when the traps undergo spontaneous
evanescence. In that case one has $Q_{1,T}>Q_1$ (see next section).

In order to compute $Q_1(\mathbf{r}_0,t;R)$, we must first define the relevant boundary value problem
by complementing Eq.~\eqref{fde} with the deterministic initial condition,
\begin{equation}
w(\mathbf{r},0|\mathbf{r}_0;0)=\delta(\mathbf{r}-\mathbf{r}_0),
\end{equation}
and the boundary conditions
\begin{align}
\label{bc1}
w(R,t|\mathbf{r}_0;0) &=0, \\
\lim_{r\to\infty} w(\mathbf{r},t|\mathbf{r}_0;0) &=0 .
\end{align}
The boundary condition \eqref{bc1} reflects the fully absorbing nature of the target,
which prevents the trap from being found on the target surface or inside the target.
The solution $w(\mathbf{r},t|\mathbf{r}_0;0)$ is related to $Q_1(\mathbf{r}_0,t;R)$ via the integral relation
\begin{equation}
\label{Q1w}
 Q_1(\mathbf{r}_0,t;R)=\int  w(\mathbf{r},t|\mathbf{r}_0;0)\, d\mathbf{r}.
\end{equation}
The spherical symmetry of the target means that $Q_1(\mathbf{r}_0,t;R)$ only depends on the initial distance $r_0$ of the trap to the target.  For this reason, we shall drop the subindex of $r_0$ and from here on use the simpler notation $Q_1(r,t;R)$. Taking into account Eq. \eqref{Q1w}, the boundary value problem stated directly in terms of $Q_1(r,t;R)$ then is
\begin{subequations}
\label{bvpQ1}
\begin{align}
\frac{\partial Q_1({r},t;R)}{\partial t} & =K_{\gamma}
 \,~_{0} D_t^{1-\gamma} \, \nabla_{\mathbf{r}}^2\,Q_1({r},t;R) \\
Q_1({r},0;R) &=1, \label{incondQ1} \\
Q_1( R,t;R) &=0, \label{bcQ1a} \\
\lim_{{r}\to\infty} Q_1({r},t;R) &=1. \label{bcQ1b}
\end{align}
\end{subequations}
The second equation in this set
corresponds to the initial condition and is self-explanatory, while
the third one is a boundary condition which reflects yet again the fully
absorbing nature of the target. The last equation states that
a trap which is ``pushed'' infinitely far away from the target will survive forever,
as its only decay channel is provided by the interaction with the target.

The above problem can be solved exactly in Laplace space~\cite{targetyuste},
\begin{equation}
\label{lapQ1}
   u {Q}_1(r,u;R)=
   1-\left(\frac{r}{R}\right)^{1-\frac{d}{2}}
   \frac{K_{d/2-1}\left(\sqrt{r^2
u^\gamma/K_\gamma}\right)}{K_{d/2-1}\left(\sqrt{R^2 u^\gamma/K_\gamma}\right)},
 \end{equation}
where $K_{d/2-1}(\cdot)$ is a modified Bessel function of the second kind.
Alternatively, the solution ${Q}_1(r,u;R|\gamma)$ for $\gamma<1$ can
be found from the corresponding solution for normal diffusion ($\gamma=1$)
by means of the ``time-expanding transformation"~\cite{SokoTET,LaplaceScaling} associated
with the so-called subordination principle: $uQ_1(r,u;R|\gamma)=u^{\gamma} Q_1(r,u^\gamma;R|\gamma=1)$.

For $d=1$ and $d=3$ the Bessel functions can be expressed in terms of exponentials,
and explicit exact solutions are available for arbitrary times $t$. In other dimensions simple
expressions are only available at long times.
Setting $Q_{1,T}=Q_1$ in Eq.~\eqref{Q*Ta}, Tauberian theorems can be used to find
the long-time behavior in the multiple trap problem~\cite{targetyuste}:
\begin{equation}
\label{genbeh}
\sigma(t;R)\propto
\left\{
\begin{array}{c l}
t^{\gamma/2} \qquad& d=1 \\
t^{\gamma}/\ln(\alpha_\gamma t) \qquad& d=2,\\
t^{\gamma} \qquad &d\ge 3
\end{array}\right.,
\end{equation}
where $\alpha_\gamma=(4K_\gamma/R^2)^{1/\gamma}$.

Thus, the survival probability of the target goes to zero in all
dimensions $d$. This result is in strong contrast  with the single-trap problem,
since in that case the probability that the random walk of the trap never intersects
the target becomes non-zero as soon as $d\ge 3$.

\section{Survival probability for evanescent traps: general expressions}
\label{genexpressions}

The behavior of the survival probability of the target changes completely if the
traps disappear in the course of their motion.
We assume a spontaneous evanescence process, specifically, that
the decrease of the global trap density $\rho(t)$ is described by the
following differential equation:
\begin{equation}
\label{nonconstlamb}
\dot\rho(t)\equiv\frac{d \rho(t)}{dt}=-\lambda(t)\, \rho(t),
\end{equation}
where $\lambda(t)>0$ is a rate coefficient which is in general time dependent.
The solution $\rho_0\,\exp{\left(-\int_0^t \lambda(t')\,dt'\right)}$ yields a
decaying density of surviving traps which describes the time evolution
of the trap density in the absence of the fully absorbing target. The
case $\lambda=$ constant leads to an exponentially decaying density.

Our main goal is to compute the survival probability $Q^*_T(t;R)$ of the target at
time $t$ (we use survival probabilities with a star to distinguish these quantities
from their counterparts in the absence of trap evanescence). We
follow the strategy of the previous section, namely, to derive the solution from
the single-trap case.

We wish to combine the effects of Eq.~\eqref{nonconstlamb} and Eq.~\eqref{fde}.
One might be tempted to proceed as in the case of
ordinary diffusion and simply construct some superposition of transport and reaction
terms. However, a careful analysis shows that this is
\textit{incorrect}.  Instead, a rigorous derivation starting at the level of the CTRW
shows that the correct equation is~\cite{AbadYusteLindenberg10}
\begin{eqnarray}
\frac{\partial w(\mathbf{r},t|\mathbf{r}_0,0)}
{\partial t}&=&\frac{\rho(t)}{\rho_0}\, K_{\gamma}
~_{0}{D}_t^{1-\gamma} \frac{\rho_0}{\rho(t)} \nabla_{\mathbf{r}}^2
w(\mathbf{r},t|\mathbf{r}_0,0)\nonumber\\
&&+\frac{\dot{\rho}(t)}{\rho(t)} w(\mathbf{r},t|\mathbf{r}_0,0).
\label{RSE}
\end{eqnarray}
It is straightforward to show that the survival probability
$Q^*_1(\mathbf{r}_0,t;R)=Q^*_1(r_0,t;R)=\int w(\mathbf{r},t|\mathbf{r}_0;0)\, d\mathbf{r}$ of the trap then
obeys the equation
\begin{subequations}
\label{bvpQ1*}
\begin{widetext}
\be
\frac{\partial Q^*_1({r},t;R)}
{\partial t}=\frac{\rho(t)}{\rho_0}K_{\gamma}
~_{0}{D}_t^{1-\gamma}
\left(\frac{\rho_0}{\rho(t)}\nabla_{\mathbf{r}}^2\right)Q^*_1({r},t;R)
+ \frac{\dot \rho(t)}{\rho(t)} Q^*_1({r},t;R),
\label{geneqQ*1arbrho2}
\ee
\end{widetext}
 where we have again dropped the subindex from $r_0$ for notational convenience. 
The above equation must now be complemented with the conditions
\begin{align}
Q^*_1(r,0;R) &=1, \\
Q^*_1(R,t;R) &=0, \\
\lim_{r\to\infty} Q^*_1(r,t;R) &=\frac{\rho(t)}{\rho_0}.
\end{align}
\end{subequations}
Note that the last equation is substantially different from the corresponding one in the
absence of the evanescence process. Indeed, even the survival probability of a trap which
is at an infinite distance from the target decays in time because of the evanescence reaction,
and this probability is equal to the ratio of the global trap density at
time $t$ in the absence of the target and the initial density $\rho_0$.

At this stage, one can easily check that if one performs the transformation
$Q^*_1(r,t;R)=[\rho(t)/\rho_0] {\mathcal Q}(r,t;R)$ in Eqs.~\eqref{bvpQ1*},
the resulting set of equations for the transformed function $\mathcal Q$ is identical
with the boundary value problem~\eqref{bvpQ1} for $Q_1$. Because of the uniqueness
of the solution, we thus conclude ${\mathcal Q} \equiv Q_1$, and further,
\begin{equation}
\label{Q1*Q1}
Q^*_1(r,t;R)=\frac{\rho(t)}{\rho_0} Q_1(r,t;R).
\end{equation}
This expression is intuitively clear: it simply states that the probability that up to time
$t$ the trap has neither evanesced (one decay channel)
nor hit the target (another decay channel) is equal to the product of the
probability $\rho(t)/\rho_0$ that the trap has not evanesced, and the conditional
probability that it has not hit the target given that it has not previously evanesced.
The latter probability is precisely the survival probability of the trap when
no evanescence is at play.

The next step in our route to the solution for the multitrap problem is to derive
a relation between $Q_{1,T}^*(r,t;R)$ and $Q_1^*(r,t;R)$. As already
anticipated in the previous section, the evanescence of the trap implies that
the survival probability of the target and the trap are no longer the same. Let $t'$ be a time in the
interval $(0,t)$.
In the single-trap problem, the infinitesimal probability $\{d\,[1-Q^*_{1,T}(r,t';R)]/dt'\}dt'$
that the target is annihilated by collision with the trap during the interval $(t',t'+dt')$ is the
product of two factors: i) the probability $\{d\,[1-Q_1(r,t';R)]/dt'\}dt'$ that the trap
collides with the target during the time interval $(t',t'+dt')$ given that it has not previously
evanesced, and ii) the probability $\rho(t')/\rho_0$ that up to time $t'$ the trajectory of that trap is not
interrupted by an evanescence event. Thus,
\begin{equation}
\frac{d}{dt'}[1-Q^*_{1,T}(r,t';R)] dt'=\frac{\rho(t')}{\rho_0}\frac{d}{dt'}[1-Q_{1}(r,t';R)] dt'.
\end{equation}
We next implement a number of steps [integrate this equation, integrate by parts, and use
Eq.  (\ref{Q1*Q1})] to obtain
\begin{equation}
\label{Q*1TQ1*b}
Q^*_{1,T}(r,t;R)=Q^*_{1}(r,t;R)-\int_0^t Q^*_{1}(r,t';R)
\frac{\dot \rho(t')}{\rho(t')}\,dt',
\end{equation}
which quantifies the difference between $Q^*_{1,T}(r,t;R)$ and $Q^*_{1}(r,t;R)$.

Having clarified the relation between $Q^*_1(r,t;R)$ and $Q_1(r,t;R)$ (survival probability of a single non-evanescent trap) and between $Q^*_{1,T}(r,t;R)$ (survival probability of the target in the presence of a single evanescent trap) and $Q^*_1(r,t;R)$ (survival probability of a single evanescent trap), we are now ready to tackle the multiple trap problem by proceeding as in the previous section, i.e., by using the statistical independence of the traps. Equations~\eqref{Q*Ta} (for the survival probability of the target in the presence of a collection of traps) and \eqref{sigmatR} are now respectively replaced with
\begin{eqnarray}
  Q^*_T(t;R)&=&\lim_{N_0,V\to\infty} \left[\frac{1}{V} \int_{r>R}
  Q^*_{1,T}(r,t;R)\, d\mathbf{r} \right]^{N_0}\nonumber\\\nonumber\\
&=&\exp\left\{-\rho_0 R^d \sigma^*(t,R) \right\},
\label{Q*Tc}
\end{eqnarray}
and
\begin{eqnarray}
 \sigma^*(t;R)&=&
 \frac{1}{R^d} \int_{r>R}[1-Q^*_{1,T}(r,t;R)]\,d\mathbf{r} \label{defsigma*0}\\ \nonumber\\
 &=& -\frac{1}{\rho_0 R^d} \ln Q^*_T(t;R).
\label{defsigma*}
\end{eqnarray}
Taking the derivative of Eq.~\eqref{defsigma*0} with respect to time and using Eq.~\eqref{Q*1TQ1*b} we get
\begin{equation}
\label{sigma*tRb}
\frac{\partial \sigma^*(t;R)}{\partial t}= -\frac{1}{R^d}\int_{r>R}
\left[\frac{\partial Q^*_{1}}{\partial t}
-\frac{\dot \rho(t)}{\rho(t)} Q^*_1\right]\,d\mathbf{r}.
\end{equation}
Next we use Eq.~\eqref{geneqQ*1arbrho2} and the relation \eqref{Q1*Q1} in the right
hand side of Eq.~\eqref{sigma*tRb} and apply Gauss' theorem to change the volume
integral to a surface integral. This allows us to write
\begin{equation}
\frac{\partial \sigma^*(t;R)}{\partial t}=S_d \frac{K_{\gamma}}{R} \frac{\rho(t)}{\rho_0} ~_{0}D_{t}^{1-\gamma} \left. \frac{\partial
Q_1(r,t;R)}{\partial r}\right|_{r=R},
\end{equation}
where $S_d=2\pi^{d/2}/\Gamma(d/2)$ denotes the surface of a $d$-dimensional
hypersphere of unit radius.
Finally, integrating from $0$ to $t$ and using the condition
$\sigma^*(0;R)=0$ [$Q^*_{1,T}(0,R)=1$] we obtain the general formula
\begin{eqnarray}
\sigma^*(t;R)&=&S_d \frac{K_{\gamma}}{R} \int_0^t  \left[~_{0}D_{t'}^{1-\gamma} \left.
\frac{\partial Q_1(r,t';R)}{\partial r}\right|_{r=R}\right]\nonumber\\
&&\times \frac{\rho(t')}{\rho_0}\,dt',
\label{sigmaAlt2}
\end{eqnarray}
which extends the result for non-evanescent traps obtained in \cite{targetyuste} to
the case of evanescent traps. More specifically, we have again related the logarithm of the survival probability of the target, now in the presence of a collection of \emph{evanescent} traps, to the survival probability of the target in the presence of a single non-evanescent trap. If the
trap density $\rho(t)$ decays sufficiently rapidly, then $Q^*_{1,T}(t\to\infty,R)>0$
and $\sigma^*(t\to\infty;R)<\infty$, and it is possible to conclude from Eq.~\eqref{sigmaAlt2} that
\begin{widetext}
\begin{equation}
\label{sigmaAlt3}
\sigma^*(\infty;R)-\sigma^*(t;R)=S_d \frac{K_{\gamma}}{R} \int_t^\infty  \left[~_{0}D_{t'}^{1-\gamma} \left. \frac{\partial Q_1(r,t';R)}{\partial r}\right|_{r=R}\right] \frac{\rho(t')}{\rho_0}\,dt'.
\end{equation}
\end{widetext}
To continue from here we need to specify the dimensionality explicitly, and also the explicit
form of the trap density as a function of time.  First we discuss the dimensionality. The case $d=1$ has been dealt with in ~\cite{YRLevanescenciaPRE}; suffice it to say that we recover the result obtained
therein:
\begin{equation}
\label{oldresult}
\sigma^*(t;R)= \frac{1}{\rho_0 R}\,\frac{2\sqrt{K_{\gamma}}}{\Gamma(\gamma/2)}
\int_0^t \rho(t') \,t'^{\gamma/2-1} \, dt',
\end{equation}
where we have made use of the explicit one-dimensional form of $Q_1$ in terms of the Fox $H$-function
\be
Q_1(r,t;R)=1-H_{11}^{10}\left[\frac{r-R}{\sqrt{K_\gamma t^\gamma}}\LM{(1,\gamma/2)}{(0,1)}\right]
\ee
for the survival probability of a non-evanescent trap.
For $d=3$, and exact expression for the survival probability valid for arbitrary $t$ can
also be obtained. This is a new result so we proceed in more detail.
For the single-trap problem without evanescence we have
~\cite{targetyuste}
\begin{equation}
Q_1(r,t;R)=1-\frac{R}{r}H_{11}^{10}\left[\frac{r-R}{\sqrt{K_\gamma t^\gamma}}
\LM{(1,\gamma/2)}{(0,1)}\right],
\end{equation}
leading to
\begin{equation}
\left. \frac{\partial
Q_1(r,t;R)}{\partial r}\right|_{r=R}
= \frac{1}{R}+
\frac{1}{\Gamma(1-\gamma/2)}\,\frac{1}{\sqrt{K_\gamma t^\gamma}}.
\end{equation}
With this result we get from Eq.~\eqref{sigmaAlt2}
\begin{eqnarray}
\sigma^*(t;R)&=&\frac{4\pi \sqrt{K_{\gamma}}}{ \rho_0 \Gamma(\gamma/2)}
\int_0^t \rho(t') \,t'^{\gamma/2-1}\,dt'\nonumber\\
&&+\frac{4\pi K_{\gamma}}
{\rho_0 R \Gamma(\gamma)}\int_0^t \rho(t') \,t'^{\gamma-1}\,dt'.
\label{sigma3d}
\end{eqnarray}
Here we have used the fact that the fractional derivative
of $C$, a constant, is not zero but is instead
$~_{0}D_{t}^{1-\gamma} C= C t^{\gamma-1}/ \Gamma(\gamma)$.

In contrast with the $d=1$ and the $d=3$ cases, no explicit solution in a simple integral form similar to that of Eqs.~\eqref{oldresult} and \eqref{sigma3d} is available for
$\sigma^*(t;R)$ when $d=2$. However, we can write explicit
expressions for the approach of the survival probability to its final value. From the
result for $Q_1$ given in ~\cite{targetyuste}, one finds
\begin{equation}
\left.\frac{\partial Q_1(r,t;R)}{\partial r}\right|_{r=R} \sim \frac{2}{R}
\frac{1}{\gamma \ln(\alpha_\gamma t)}.
\label{derQ12d}
\end{equation}
The fractional derivative of this expression is best computed
in Laplace space~\cite{targetyuste}. Transforming the resulting expression back into the
time domain we find that
\be
~_{0}D_{t}^{1-\gamma}\left[1/\ln(\alpha_\gamma t)
\right]\sim t^{\gamma-1}/[\Gamma(\gamma)\ln(\alpha_\gamma t)].
\label{derlog}
\ee
Let us further assume that the trap density $\rho(t)$ decays sufficiently rapidly to ensure
that $\sigma^*(\infty;R)$ is finite. From Eq.~\eqref{sigmaAlt3} we find that for $d=2$,
\begin{equation}
\label{asmu2}
\sigma^*(\infty;R)-\sigma^*(t;R)
\sim \frac{4\pi K_\gamma}{\rho_0 R^2\Gamma(\gamma+1)}\int_t^\infty
\frac{\rho(t')\, t'^{\gamma-1}}{\ln(\alpha_\gamma t')}\,dt'.
\end{equation}


We next implement our general results for particular forms of the decay of the trap density.

\section{Implementation for particular trap decay functions}
\label{implementation}

\subsection{Exponentially decaying trap density}

We first consider an exponentially decaying trap density,
$\rho(t) = \rho_0\exp(-\lambda t)$, where the characteristic time scale of the decay is given by
$\tau\equiv\lambda^{-1}$. This would represent a unimolecular decay
if this were the only decay channel, which in the presence of the target it is not.  However, for a
single particle, or for the first of many, this is still the scenario because we are looking at the
decay only up to the time that this second channel first affects the trap density.
The focus of our interest is in the final value of the survival probability of the target and the
long time approach to it, but for completeness we also give a general expression for the early time behavior.

For $d=1$ and $d=3$ we can directly insert the exponential decay of the density in
Eqs.~(\ref{oldresult}) and (\ref{sigma3d}), and perform the integrals.
Alternatively, we note that for the exponentially decaying trap density Eq.~(\ref{Q*1TQ1*b}) leads to the relation
\begin{eqnarray}
Q_{1,T}^*(r,t;R) &=& e^{-\lambda t} Q_1(r,t;R)
\nonumber\\
&&+\lambda \int_0^t e^{-\lambda t^\prime} Q_1(r,t^\prime; R) dt^\prime.
\label{gettingthere}
\end{eqnarray}
Next, from Eq.~(\ref{lapQ1}) and the Laplace transform of Eq.~(\ref{gettingthere}) it then follows that
\begin{widetext}
\begin{equation}
\label{Q*1TK}
u {Q}^*_{1,T}(r,u;R)=
1- \left(\frac{r}{R}\right)^{1-(d/2)}\,
\frac{K_{d/2-1}(\sqrt{r^2(u+\lambda)^\gamma/K_\gamma})}
{K_{d/2-1}(\sqrt{R^2(u+\lambda)^\gamma/K_\gamma})}.
\end{equation}
\end{widetext}
The Laplace transform of Eq.~(\ref{defsigma*0}) then immediately leads to
\begin{equation}
\sigma^*(u;R) =\frac{S_d K_\gamma^{1/2}}{uR (u+\lambda)^{\gamma/2}} \frac{K_{d/2}(\sqrt{R^2(u+\lambda)^{\gamma}/K_\gamma})}{K_{d/2-1}(\sqrt{R^2(u+\lambda)^{\gamma}/K_\gamma})}.
\label{Laptransigma}
\end{equation}
For odd-valued $d$ (but not for even-valued $d$) the modified Bessel functions of the second kind can be expressed more simply in standard power series expansions.
For $d=1$ we can follow either path (direct integration or simplification and inversion of the modified Bessel function) to obtain
\begin{equation}
\sigma^*(t;R) = \frac{2\ell_\gamma}{R} \left( 1-\frac{\Gamma(\gamma/2,\lambda t)}{\Gamma(\gamma/2)}\right),
\label{theintegral}
\end{equation}
where $\ell_\gamma \equiv (K_\gamma \tau^{\gamma})^{1/2}$ is a characteristic length scale associated with the distance covered by a trap during its mean survival time when no target is present. Eq.~\eqref{theintegral}
is equivalent to Eq.~(15) in Ref.~\cite{YRLevanescenciaPRE}. For $d=3$ we can again follow either route to the
solution and find
\begin{eqnarray}
\sigma^*(t;R) &=& 4\pi\frac{\ell_\gamma}{R} \left( 1-\frac{\Gamma(\gamma/2,\lambda t)}{\Gamma(\gamma/2)}\right) \nonumber\\
&& +4\pi \frac{\ell_\gamma^2}{R^2} \left( 1-\frac{\Gamma(\gamma,\lambda t)}{\Gamma(\gamma)}\right),
\label{the3dintegral}
\end{eqnarray}
resulting in a smaller survival probability than in $d=1$.  As noted already, there are no simple, closed-form solutions valid for arbitrary times for $d=2$, although we are able to extract some limiting behaviors for this case as well (see below).

The computation of the final value of $\sigma^*(t;R)$ in arbitrary integer dimension
is readily obtained from ~\eqref{Laptransigma} by means of the final value theorem for the Laplace
transform:
\begin{equation}
\sigma^*(\infty;R) = \lim_{u\to 0} u\,\sigma^*(u;R) =S_d \frac{\ell_\gamma}{R}\frac{K_{d/2}(R/\ell_\gamma)}{K_{d/2-1}(R/\ell_\gamma)}.
\label{Besselfunctionway}
\end{equation}
Before further evaluation, we note as an aside that the above non-zero survival probability implies an infinite mean survival time of the target in any dimension $d$. In contrast, if the traps do not evanesce, the mean lifetime of the target is finite~\cite{Grebenkov}. For odd dimensions,   Eq.~\eqref{Besselfunctionway} can be conveniently rewritten as  rational functions of the argument $R/\ell_\gamma$. For example : $\sigma^*(\infty;R) =  2\ell_\gamma/R$ for $d=1$ and $\sigma^*(\infty;R)= (4\pi \ell_\gamma/R) \left(1+\ell_\gamma/R\right)$ for $d=3$.

Next we explicitly present the results for the approach to the final value by exhibiting the difference $\sigma^*(\infty;R)-\sigma^*(t;R)$ at long times.  For $d=1$ and $d=3$ this respectively follows from Eqs.~(\ref{theintegral}) and (\ref{the3dintegral}), whereas from $d=2$ the long time behavior can
be inferred from  Eq.~\eqref{asmu2}.  We find:
\begin{equation}
\sigma^*(\infty;R)-\sigma^*(t;R)\propto \left\{
\begin{array}{c l}
t^{\gamma /2-1}e^{-\lambda t} \qquad& d=1\\ \\
\label{aslawev}
\ln^{-1}(\alpha_\gamma t)\,t^{\gamma-1} e^{-\lambda t} \qquad &d=2\\ \\
t^{\gamma-1} e^{-\lambda t} \qquad &d=3.
\end{array}\right.
\end{equation}
 This asymptotic behavior also holds in the case of normal diffusion
 ($\gamma=1$).  One can see that the decay of the survival probability to the final state
 prescribed by Eq. \eqref{aslawev} becomes faster as one goes from one to two dimensions and
 from two to three dimensions.  It is also straightforward to show that the long time
 behavior prescribed by \eqref{aslawev} for $d=3$ remains valid for $d>3$ (the prefactor, however, depends on $d$).

 The limit $\lambda\to 0$ (no evanescence) turns out to be singular. Indeed, in the absence of evanescence, $\sigma^*(t;R)$ tends to infinity as given by Eq.~\eqref{genbeh},
which is different from the result obtained when taking the limit $\lambda\to 0$ in Eq.~\eqref{aslawev}. We conclude that the evanescence reaction completely changes the physics of the problem, affecting both the steady state and the decay form of the survival probability.

Finally, the short time behavior ($t\ll \lambda^{-1}$) is straightforward to obtain via a Tauberian theorem applied to the large $u$ ($u\gg \lambda$) limit of Eq.~(\ref{Laptransigma}):
\begin{equation}
\sigma^*(u;R) \sim \frac{S_d K_\gamma^{1/2}}{Ru^{1+\gamma/2}} \to \sigma^*(t;R) \sim \frac{S_d K_\gamma^{1/2}}{R\Gamma(1+\gamma/2)} t^{\gamma/2}.
\end{equation}
As one might have guessed, the short time result is independent of $\lambda$, i.e., the effect of the evanescence
reaction is still negligible in this regime.

\subsection{Power law decay of the trap density}

We next turn to the case of a power-law decaying density, that is,
\begin{equation}\label{powerlaw}
\rho(t)=\frac{\rho_0}{(1+t/\tau)^\beta}, \qquad \beta>0.
\end{equation}
This choice corresponds to a time dependent rate constant $\lambda(t)$, which can be used to
capture the essential features of complex higher-order kinetics by means of the linear differential equation
\eqref{nonconstlamb} and a proper choice of the exponent $\beta$. Interestingly,  the survival probability of the target in this case depends not only on dimensionality but on the relative values of the power law decay exponent $\beta$ and the subdiffusion exponent $\gamma$.

The behavior when $d=1$ follows directly from Eq. \eqref{oldresult}, and can be summarized as follows
\cite{YRLevanescenciaPRE}:
\begin{equation}
\label{oldresult2}
\sigma^*(t;R) \sim \begin{cases}
\ell_\gamma \frac{\Gamma(\beta-\gamma/2)}{R\Gamma(\beta)},  \qquad &\beta>\gamma/2 \\
\frac{\ell_\gamma}{R\Gamma(\gamma/2)}\ln (t/\tau),  \qquad &\beta=\gamma/2 \\
 \frac{\ell_\gamma}{R(\gamma/2-\beta)\Gamma(\gamma/2)}(t/\tau)^{\gamma/2-\beta}, \quad &\beta<\gamma/2 \end{cases}
\end{equation}
where $\ell_\gamma$ is defined in a way similar to the exponential case, i.e., $\ell_\gamma=(K_\gamma \tau^\gamma)^{1/2}$.

Next we consider the two-dimensional system. Our starting equation is \eqref{sigmaAlt2} with $d=2$:
\begin{eqnarray}
\label{sigmaAlt2d=2}
\sigma^*(t;R)&=&2\pi \frac{K_{\gamma}}{R} \int_0^t
\left[~_{0}D_{t'}^{1-\gamma} \left.
\frac{\partial Q_1(r,t';R)}{\partial r}\right|_{r=R}\right]\nonumber \\
&&\times (1-t'/\tau)^{-\beta}\,dt'.
\end{eqnarray}
The behavior of the integral on the right hand side depends on the
relative values of $\beta$ and $\gamma$. We consider three different cases:

{\bf Case 1 ($\beta>\gamma$).} From the asymptotic long-time behavior \eqref{derQ12d}
and the expression for the fractional derivative of the inverse logarithm \eqref{derlog}
one finds
\be
\label{approxlarget}
~_{0}D_{t'}^{1-\gamma} \left.
\frac{\partial Q_1(r,t';R)}{\partial r}\right|_{r=R}
\sim \frac{2\,t'^{\gamma-1}}{R \Gamma(\gamma+1)\ln(\alpha_\gamma t')}.
\ee
Under the assumption that $t$ is large enough, we now split the
interval of integration $[0,t]$
into two subintervals, $I_1=[0,t_c]$ and $I_2=[t_c,t)$, where $t_c$ is
chosen sufficiently large so as to ensure that the approximation
\eqref{approxlarget} holds over the full extent of $I_2$.
Hence one has
\be
\label{asbehsigma}
\sigma^*(t;R)\sim {\cal C}+\frac{4\pi K_\gamma \tau^\beta}{\Gamma(\gamma+1)\,R^2}
\int_{t_c}^t  \frac{t'^{\gamma-\beta-1}}{\ln(\alpha_\gamma t')}\,dt',
\ee
where we have used the long-time approximation $\rho(t')/\rho_0\approx
(t'/\tau)^{-\beta}$ and ${\cal C}$ represents the integral from $0$ to $t_c$.
In this case one can easily check via partial integration
that the integral on the right hand side of Eq.~\eqref{asbehsigma} remains finite
as $t\to\infty$. Hence $\sigma^*(\infty;R)$
is finite and the target has a non-zero chance of eternal survival.
Using the explicit form of $\rho(t')$ in Eq. \eqref{asmu2} we find
\begin{align}
\sigma^*(\infty;R)-\sigma^*(t;R) &\sim \frac{4\pi K_\gamma
\tau^\beta}{(\beta-\gamma)\Gamma(\gamma+1)\,R^2}
\frac{t^{\gamma-\beta}}{\ln(\alpha_\gamma t)}.
\end{align}
Unfortunately, it does not seem possible to find an explicit exact expression for
$\sigma^*(\infty;R)$ due to the lack of an exact expression for $Q_1(r,t';R)$ valid for
the whole time domain.

{\bf Case 2  ($\beta=\gamma$).}
In this marginal case the target also disappears eventually, but the
approach to the empty state has a different analytic
dependence, as Eq.~\eqref{asbehsigma} now leads to
\be
\sigma^*(t;R)\sim \frac{4\pi K_\gamma
\tau^\gamma}{\Gamma(\gamma+1)R^2}\ln (\ln(\alpha_\gamma t)).
\ee
Thus, the target is eventually killed with certainty, in agreement
with the result given in Ref.~\cite{Brayetal} for the special case $\beta=\gamma=1$ (normal
diffusive traps).

{\bf Case 3 ($\beta<\gamma$).}
Clearly, in this case the constant ${\cal C}$  of Eq.~\eqref{asbehsigma} becomes negligible at sufficiently long times and the
behavior of $\sigma^*(t;R)$ is
dominated by the integral on the right hand side. Using
partial integration one
easily sees that the leading long-time behavior is given by
\be
\sigma^*(t;R)\sim \frac{4\pi K_\gamma
\tau^\beta}{(\gamma-\beta)\Gamma(\gamma+1)\,R^2}
\frac{t^{\gamma-\beta}}{\ln(\alpha_\gamma t)}.
\ee
Hence when $\beta < \gamma$ the target eventually disappears with certainty.

We next discuss the three-dimensional case. In this
case, the integrals in Eq.~\eqref{sigma3d} can be carried out exactly, and one finds
\begin{eqnarray}
\label{sigmaincbeta}
\sigma^*(t;R)&=&\frac{2\pi \ell_\gamma}{\Gamma(\gamma/2)}B_{t/(\tau+t)}(\gamma/2,\beta-\gamma/2)\nonumber\\ \nonumber\\
&&+\frac{\pi \ell_\gamma^2}{R\,\Gamma(\gamma)}
B_{t/(\tau+t)}(\gamma,\beta-\gamma),
\end{eqnarray}
where
\begin{equation}
B_\alpha(z,w) = \int_0^\alpha dt \; t^{z-1} (1-t)^{w-1} \quad
\text{with } \mbox{Re}(z)>0.
\label{ibf}
\end{equation}
is the incomplete Beta function \cite{AbramowitzStegun}. The long-time behavior of
$\sigma^*(t;R)$ again depends on the relative values of $\beta$ and $\gamma$. The analysis is carried out along lines similar to those presented in Ref. \cite{YRLevanescenciaPRE} for $d=1$. We shall distinguish three different cases.

{\bf Case 1 ($\beta > \gamma$).} We can rewrite $\sigma^*(t;R)$ as
\begin{widetext}
\begin{equation}
\sigma^*(t;R)=\frac{2\pi \ell_\gamma}{\Gamma(\gamma/2)}\,B(\gamma/2,\beta-\gamma/2)
I_{t/(\tau+t)}(\gamma/2,\beta-\gamma/2)
+\frac{\pi \ell_\gamma^2}{R\,\Gamma(\gamma)}B(\gamma,\beta-\gamma)
I_{t/(\tau+t)}(\gamma,\beta-\gamma).
\end{equation}
\end{widetext}
Here $B(z,w)$ is the Beta function (where the requirement $\mbox{Re}(z)>0$
and $\mbox{Re}(w)>0$ places us in the ``Case 1" regime), and
$I_{x}(z,w)$ is the regularized incomplete Beta function as
defined in Sec.~6.6.2 (pg.~263) of Ref.~\cite{AbramowitzStegun}. Using the
property~6.6.3 in \cite{AbramowitzStegun} we can set
$I_{x}(a,b)=1-I_{1-x}(b,a)$. Applying the
relation 26.5.5 in \cite{AbramowitzStegun}, and making use of the relation
between the Beta function and the Gamma function, we arrive at the asymptotic result
\begin{equation}
\sigma^*(t;R)\sim  \sigma^*(\infty;R)-\frac{\pi \ell_\gamma^2}{R}
\frac{(t/\tau)^{\gamma-\beta}}{(\beta-\gamma)\Gamma(\gamma)}
\end{equation}
with
\begin{equation}
\sigma^*(\infty;R)=2\pi \ell_\gamma \frac{\Gamma(\beta-\gamma/2)}{\Gamma(\beta)}+\frac{\pi \ell_\gamma^2}{R}\frac{\Gamma(\beta-\gamma)}{\Gamma(\beta)},
\end{equation}
leading to a non-zero survival probability $Q_T^*(\infty;R)=\exp{[-\rho_0 R^2 \sigma^*(\infty;R)]}$.

{\bf Case 2 ($\beta =\gamma$).}  In this case the incomplete Beta function in the term proportional to $\ell_\gamma^2$
in Eq. \eqref{sigmaincbeta} can be rewritten as a hypergeometric function and consequently for long times
this term can be approximated by
\begin{equation}
\frac{\pi \ell_\gamma^2}{R \Gamma(\gamma+1)}\,
\left(\frac{t}{\tau}\right)^\gamma \; _2 F_1(\gamma,\gamma,\gamma+1,-t/\tau)
\sim \frac{\pi \ell_\gamma^2}{R\Gamma(\gamma)}\ln{(t/\tau)}.
\end{equation}
On the other hand, the term proportional to $\ell_\gamma$ goes to a constant for long times, as can be seen using the
same expansion as the one used for the $\beta>\gamma$ case. Hence, the survival probability $Q_T^*(t;R)$ vanishes as $(t/\tau)^{-\pi \rho_0 R \ell_\gamma^2/\Gamma(\gamma)}$, that is, $\sigma^*(t;R)\propto \ln (t/\tau)$.

{\bf Case 3 ($\beta<\gamma$).} In this case the term proportional to $\ell_\gamma^2$ can easily be seen to behave as
$[\pi \ell_\gamma^2/R(\gamma-\beta) \Gamma(\gamma)](t/\tau)^{\gamma-\beta}$ by performing a straightforward asymptotic analysis of the corresponding integral. On the other hand the $\ell_\gamma$ term is negligible compared to the $\ell_\gamma^2$ term. This results in a stretched exponential decay to zero, i.e., $Q_T^*(t;R)\propto \exp{(-C t^{\gamma-\beta})}$ with $C>0$, that is,  $\sigma^*(t;R) \propto t^{\gamma-\beta}$.

Thus, in two and three dimensions the target has a finite probability of surviving forever only for $\beta>\gamma$.
For comparison, in the one-dimensional case it was found that the target has a chance of eternal survival only
when $\beta>\gamma/2$ [cf. Eq. \eqref{oldresult2}].  We thus see that the interplay between subdiffusive 
transport and the evanescence reaction determines, also in dimensions higher than one, whether the target can ultimately survive. 

\section{Summary and Outlook}
\label{summary}

We have presented a particular application of a recently derived fractional reaction-subdiffusion equation, namely, the study of the behavior of the survival probability of an immobile target surrounded by a sea of
noninteracting diffusive or subdiffusive point traps subject to an evanescence reaction. The evanescence reaction is assumed to take place independently of the CTRW jumps performed by the traps, as opposed to a recently introduced model where disappearance takes place at the time of each jump \cite{Shkilev}.

The problem considered in this paper is only one of a family of many possible boundary value problems which may be dealt with using our equations. However, this particular choice may be of interest in a number of experimental situations, e.g., radical recombination kinetics in the presence of added scavenger molecules \cite{KimEtAl} also responsible for the disappearance of radicals. As far as we know, the interplay between the scavenging reaction and possible memory effects arising in some environments remains unexplored.

We focused on the case of exponential evanescence and power law evanescence,  extending previous results applicable only to the one-dimensional case. In particular, our results also hold for the normal diffusion case ($\gamma=1$). The presence of the evanescence reaction was found to completely modify the physics of the problem, both at the level of the steady state and the decay of the survival probability to a finite steady state or to zero.
More specifically, with an exponentially decaying trap density $\rho(t)=\rho_0\,e^{-\lambda t}$ (with $\lambda>0$),
we find that there is a finite survival probability of the target in all dimensions because the traps die sufficiently quickly in their search of the target. By way of contrast, when the traps do not evanesce the target has a zero survival probability in all dimensions. The long-time approach toward the final value of the
survival probability turns out to be more complex than in the case of non-evanescent traps,
and in the subdiffusive case $\gamma<1$ it involves powers of $t$ as well as exponential factors $e^{-\lambda\,t}$ (with a logarithmic correction in $d=2$). On the other hand, when the density decays as a power
law, $\rho(t)\propto t^{-\beta}$ with $\beta>0$, the behavior depends on the relative values
of $\beta$ and the anomalous diffusion exponent $\gamma$ of the traps. In one dimension,
the target has a finite asymptotic survival probability if $\beta>\gamma/2$, whereas in two and three dimensions the target only has a finite chance of eternal survival when $\beta>\gamma$.

A natural extension of this work would allow normal diffusive or subdiffusive target motion (the case
of normal diffusive target and normal diffusive evanescent traps has been considered in Ref.~\cite{DenHollanderShuler}). Note, however, that in such a case the respective distances between the target and the traps would no longer evolve as independent variables, implying that our asymptotically exact approach would not work in its present form. Nonetheless, approximations based on the fact that at long times the dominant contribution to the survival probability comes from the subset of trajectories where the target remains immobile ~\cite{BrayBlythePRL02, OshaninEtAlPRE} could prove useful to tackle the problem. Ultimately, this behavior finds its roots in what has been termed the ``Pascal principle'' in the literature~\cite{Moreau, YusteKatjaPRE05,ChenSun}, i.e., a target placed in a symmetric initial distribution of traps survives longer on average if it stays still rather than if it moves.

\begin{acknowledgments}
This work was partially funded by
the Ministerio de Ciencia y Tecnolog\'ia (Spain) through Grant No. FIS2010-16587 (partially financed by FEDER funds), by the Junta de Extremadura through Grant No. GRU10158, and by the US National Science Foundation under Grant No. PHY-0855471.
\end{acknowledgments}

\end{document}